\documentstyle[12pt]{article}
\begin{document}
\newcommand {\cD}{{\cal D}}
SPbU-IP-94

\centerline{\bf Comment on gauge choices
and physical variables in QED}

\vspace{5mm}

\centerline {Dmitri Vassilevich}

Department of Theoretical Physics, St.Petersburg
University,

198904 St.Petersburg, Russia

\vspace{5mm}

\centerline{Abstract}

We consider possible definitions of physical variables
in QED. We demonstrate that the condition $\partial_i
A_i$$=0$ is the most convenient one because it leads to
path integral over physical components with local action.
However, other choices, as $A_3=0$, are also possible.
The standard expression for configuration space path
integral in $A_3=0$ gauge is obtained starting with
reduced phase space formulation. Contrary to the claims
of the paper [1] the $A_3=0$ gauge is not overconstrained.


\vspace{5mm}

In their recent paper [1] Lavelle and McMullan claimed that
the gauge $A_3=0$ is overconstrained. The way of reasoning was
as follows. The component $A_0$ is the Lagrange multiplier and
hence non-dynamical. The physical components of the three-vector
$A_i$ are transversal. When one also imposes the condition
$A_3=0$, one can not obtain desired two helicity components
of the photon for three-momenta orthogonal to $x_3$. Hence the
gauge $A_3=0$ is overconstrained.

The above statement is obviously a result of misuse of
terminology. The notion of the Lagrange multiplier refers
to the Hamiltnian formulation of a theory, while gauge
conditions are imposed on a configuration space. The so-called
"physical components" are simply the canonical coordinates
on the reduced phase space. In QED the physical components are
fixed up to gauge transformations. It is senseless to impose
all the restrictions appearing in various formulations
simultaneously. In principle, these arguments together with
a reference to standard textbook on gauge theories [2]
(for non-covariant gauges see e.g. [3]) are enough to restore
reputation of the $A_3=0$ gauge. However, we shall do a bit more.

In this short letter we consider possible choices of physical
variables in QED. We demonstrate, that the condition
$\partial_i A_i=0$ is the most convenient one because it leads
to local action in the path integral over physical variables.
We show that it is also possible to consider the components
$A_1$ and $A_2$ as physical ones. This latter choice leads to
the standard configuration space path integral in the axial
gauge $A_3=0$. To the best of our knowledge this construction
was not considered in the literature.

The work is organized as follows. First we consider the reduced
phase space formulation of QED and comment on transition to
interactive theory. Next we derive the path integral in the
$A_3=0$ gauge.

Consider the action for electrodynamics
$$S=-\frac 14 \int d^4x F_{\mu \nu}F^{\mu \nu},
\quad F_{\mu \nu}=\partial_\mu A_\nu-
\partial_\nu A_\mu \eqno (1)$$
We can rewrite it in the first-order form
$$S=-\frac 12 \int d^4x F^{\mu \nu}
(\partial_\mu A_\nu -\partial_\nu A_\mu
-\frac 12 F_{\mu \nu})=$$
$$=\int d^4x [E_k\partial_0 A_k-\frac 12
(E^2+B^2)+A_0\Phi ] \eqno (2)$$
where $A_k$ are canonical variables and $E_k$$=F_{0k}$
are the conjugate momenta, $B_k=\frac 12 \epsilon_{kij}
F^{ij}$. The Lagrange multiplier $A_0$ generates the
constraint
$$\Phi =\partial_k E_k=0 \eqno (3)$$
which selects transversal components $E^T_k$ of $E_k$.
The first class constraint $\Phi$ in tern generates gauge
transformations. To obtain the reduced phase space one should
also impose a condition on canonical coordinates $A_k$
$$\chi (A)=0 \eqno (4)$$
The only restriction on $\chi$ is that [2,4]
$$\det \{\Phi ,\chi \} \ne 0 \eqno (5)$$
where $\{ \ ,\ \}$ are the canonical Poisson brackets.
The eq. (5) means that the condition (4) removes all the
gauge freedom related to gradient transformations of
$A_k$. We will consider only linear conditions (4).
The solutions $A^{ph}$ of eq. (4) are called physical components.
One can express the path integral in terms of $A^{ph}$
only without contribution from ghosts. According to
the general prescription [2,4,5] the evolution
operator is
$$Z=\int \cD E^T \cD A^{ph} \exp iS(E^T,A^{ph})
\eqno (6)$$
The path integral over $E^T$ is Gaussian. After integration
over $E^T$ one obtains
$$Z=\int \cD A^{ph} \exp (\frac i2 \int d^4x ((\partial_0
P^T A^{ph}_k)^2-B_k(A^{ph})^2) \eqno (7)$$
where the $P^T$ is the transversal projector,
$$P^T=\delta^k_i-^{(3)}\Delta^{-1}\partial_i
\partial^k, \eqno (8)$$
$\ ^{(3)}\Delta$ is the three-Laplacian. We see, that the
expression
in the exponential of (7) is local if and only if $\chi$ is the
tarnsversality condition
$$\chi (A)=\chi^T(A)=\partial_i A_i \eqno (9)$$
The choice (9) is obviously the most convenient one, but it is
not unique.

Transition to interactive theory is made by introducing the
term $A_\mu J^\mu$ in the action (1). The constraint (3) is
replaced by
$$\Phi +J_0 =0 \eqno (10)$$
If $\chi$ depends only on $A$, the Poisson bracket (5) do not
change because $J_0$ do not depend on canonical momenta $E^k$
$$\{ \chi (A),\Phi +J_0\} =\{ \chi ,\Phi \} \eqno (11)$$
and all our arguments are still valid.

One can extend the integration region in (6)
$$Z=\int \cD E \cD A_0 \cD A_i \delta (\chi )
\det \{ \Phi ,\chi \} \exp iS(E,A_i,A_0), \eqno (12)$$
where $S$ is the first order action (2). By integrating
over $E$ we can arrive at gauge fixed path integral
$$Z=\int \cD A_\mu \delta (\chi ) \det \{ \Phi ,\chi \}
\exp iS(A_\mu) \eqno (13)$$
with the action (1).

As a condition $\chi$ one can choose $\chi_3 (A)=A_3=0$ and
$\tilde \chi $$=\partial_i A_i (x_3=0)=0$. The determinant
$\det \{ (\chi_3, \tilde \chi ), \Phi \} \ne 0$ because the
condition $\chi_3$ fixes all the gauge freedom except for
$x_3$-independent gauge transformations, and the remaining
gauge freedom is removed by the condition $\tilde \chi$$=0$
imposed on a hypersurface $x_3=$const. The necessity of the
condition $\tilde \chi$ is well known. We see that the components
$A_1$ and $A_2$ themselves can play the role of physical
variables. Hence the gauge $A_3=0$ is not overconstrained.
Though the action in (7) is non-local, the configuration
space path integral (13) has the standard form with local
action
$$Z=\int \cD A_\mu \delta (A_3) \delta (\tilde \chi )
\det \{ (\chi_3, \tilde \chi ), \Phi \}
\exp (-\frac i4 \int d^4x F_{\mu \nu}F^{\mu \nu})
\eqno (14)$$
For most of the physical applications the Jacobian factor
$\det \{ (\chi_3, \tilde \chi ), \Phi \}$ can be dropped out
as an irrelevant constant. However, for non-trivial topology
and/or geometry of space-time it should be retained. In this
latter case the path integral measure can have rather
non-trivial structure [6].

In this letter we demonstrated that the choice of physical
variables $\partial_i A_i=0$ is not unique but most convenient.
The components $A_1$ and $A_2$ (subject to the constraint
$\tilde \chi =0$ on a $x_3$=const plane) can also be chosen
as physical variables. This leads to the standard path integral
in the $A_3=0$ gauge. The result of the paper [1] that the
$A_3=0$ gauge is overconstrained is incorrect.

This work was supported by the Russian Foundation for
Fundamental Studies, grant \# 93-02-14378.

\vspace{5mm}

{\bf References}
\newline
1. M.Lavelle and D.McMullan, Phys. Lett. B316 (1993) 172.
\newline
2. L.D.Faddeev and A.A.Slavnov, Gauge fields: Introduction to
quantum theory, Benjamin/Cummings, 1980.
\newline
3. P.Gaigg, W.Kummer and M.Schweda, eds., Physical and Nonstandard
gauges, Springer, 1990.
\newline
4. L.D.Faddeev, Teor. Mat. Fiz. 1 (1969) 3.
\newline
5. L.D.Faddeev and V.N.Popov, Phys. Lett. 25B (1967) 29.
\newline
6. D.V.Vassilevich, Nuovo Cimento 105A (1992) 649;

I.P.Grigentch and D.V.Vassilevich, Nuovo Cimento 107A (1994) 227.

\end{document}